\documentclass{article}
\title{Holonomic Gradient Descent  \\ and its Application to Fisher-Bingham Integral
}
\author{Tomonari Sei, Nobuki Takayama, Akimichi Takemura ;\\
        Hiromasa Nakayama, Kenta Nishiyama, Masayuki Noro, Katsuyoshi Ohara}
\date{September 5, 2010}

\usepackage{graphicx}
\newtheorem{algorithm}{Algorithm}

\newtheorem{example}{Example}
\newtheorem{proposition}{Proposition}
\newtheorem{remark}{Remark}
\newtheorem{theorem}{Theorem}
\newtheorem{lemma}{Lemma}

\def\pd#1{\partial_{#1}}
\def\comment#1{ }
\def\QED{Q.E.D. \bigbreak }

\begin{document}
\maketitle

We give a new algorithm to find local maximum and minimum 
of a holonomic function and apply it 
for the Fisher-Bingham integral on the sphere $S^n$,
which is used in the directional statistics.
The method utilizes the theory and algorithms of holonomic systems.

\section{Introduction}

The gradient descent is a general method to find a local minimum
of a smooth function $f(z_1, \ldots, z_d)$.
The method utilizes the observation that $f(p)$ decreases if one goes
from a point $z=p$ to a ``nice'' direction,
which is usually $-(\nabla f)(p)$.
As textbooks on optimizations present
(see, e.g., \cite{iwanami}, \cite{synman}),
we have a lot of achievements on this method and its variations.

We suggest a new variation of the gradient descent,
which works for  real valued {\it holonomic functions} $f(z_1, \ldots, z_d)$
and is a $d$-variable generalization of Euler's method for
solving ordinary differential equations numerically
and finding a local minimum of the function.
We show an application of our method to directional statistics.
In fact, it is our motivating problem to develop the new method.

A function $f$ is called a holonomic function,
roughly speaking, 
if $f$ satisfies a system of linear differential equations
\begin{equation} \label{eq:system}
 \ell_1 \bullet f = \ldots = \ell_r \bullet f = 0, \quad
 \ell_i \in D
\end{equation}
whose solutions form a finite dimensional vector space.
Here, $D$ is the ring of differential operators
with polynomial coefficients
${\bf C}\langle z_1, \ldots, z_d, \partial_1, \ldots, \partial_d \rangle$,
and the action $\bullet$ is defined by
$ z^\alpha \partial^\beta \bullet f = 
 z_1^{\alpha_1} \cdots z_d^{\alpha_d} 
    \frac{ \partial^{|\beta|} f}
         {\partial z_1^{\beta_1} \cdots \partial z_d^{\beta_d}}$.

Let us give a rigorous definition of holonomic function.
A multi-valued analytic function $f$ defined on ${\bf C}^d \setminus V$
with an algebraic set $V$ is called a {\it holonomic  function}
if there exists a set of linear differential operators $\ell_i \in D$
annihilating $f$ as (\ref{eq:system})
such that the left ideal generated by 
$\{ \ell_1, \ldots, \ell_r \}$ in $D$
is a {\it holonomic ideal}
(see \cite{SST}). 
The function $f$ is called  real valued 
when a branch of $f$ takes real values on a connected component 
of $({\bf C}^d \setminus V) \cap {\bf R}^d$.

We give an equivalent definition of
holonomic function without the notion of the holonomic ideal 
( \cite{takayama1992}, \cite{OTW}, \cite{SST}).
A multi-valued analytic function $f$ is called a holonomic function
if $f$ satisfies linear ordinary differential equations
with polynomial coefficients for all variables $z_1, \ldots, z_d$.
In other words, the function $f$ satisfies a set of ordinary differential 
equations
$$ \sum_{k=0}^{r_i} a^i_k(z_1,\ldots, z_d) \partial_i^k \bullet f = 0, \quad
   a^i_k \in {\bf C}[z_1, \ldots, z_d], \quad i=1, \ldots, d.
$$
When $n=1$, a holonomic function is nothing but a solution
of linear ordinary differential equation with polynomial coefficients.
In this case, a local minimum can be obtained numerically by a difference scheme,
which is called Euler's method.
Readers may think that it will be straight forward to generalize Euler's
method to $d$-variables, which we will call {\it holonomic gradient descent}.
However, as we will see in this paper,
a generalization of Euler's method to $d$-variables requires
to utilize the theory, algorithms, and efficient implementations
of Gr\"obner basis for holonomic systems, 
which have been studied recently (see \cite{SST} and its references).

In Section \ref{sec:hgd}, we will illustrate  holonomic gradient descent precisely.
In Sections \ref{sec:fbi} and \ref{sec:hsfbi}, we study the Fisher-Bingham
integral as a holonomic function.
In Section \ref{sec:cr}, we consider problems in the directional statistics
as applications of results of Sections \ref{sec:hgd}, \ref{sec:fbi}, and \ref{sec:hsfbi}.

Our method is based on holonomic systems of differential equations.
D.~Zeilberger proposed 
the holonomic function approach for special function identities
about 20 years ago and it has been studied in the past 20 years
(see, e.g., \cite{APZ} and its references).
We present, in this paper, that the holonomic approach will be promissing
as a new method in statistics and in optimization.
We note that this point of view of holonomic systems
and holonomic functions
has been emphasized by few literatures in statistics and in optimization.

\section{Gradient Descent for Holonomic Functions}  \label{sec:hgd}

There are several methods  of finding a local
minimum of a given function $g$.
Among them, iteration methods are the most general and are often used
methods.
Iterations are written as
\begin{equation}
 z^{(k+1)} = z^{(k)} + h_k d^{(k)}  \quad k=0,1,2, \ldots
\end{equation}
where $\{ z^{(k)} \in {\bf R}^d \}$ is a sequence such that
$g(z^{(k)})$ converges to a local minimum of the function $g$,
$h_k \in {\bf R}_{>0}$ is a step length,
and $d^{(k)}$ is called the search direction.
The search direction has the form
\begin{equation}  \label{eq:search-direction}
 -H_k^{-1} (\nabla g)(z^{(k)}) 
\end{equation}
where $H_k^{-1}$ is a $d \times d$ matrix.
Typical choices of $H_k$ are the identity matrix for the gradient descent
and the Hessian matrix of $g$ for Newton's method
\cite{iwanami}.

The iteration method is a numerical method.
When the function $g$ is a holonomic function, we can apply 
the Gr\"obner basis method, which is an algebraic and symbolic method, 
for the evaluation of the search direction.
When we are given a Gr\"obner basis $B$, a set of monomials $S$ is called
the set of the {\it standard monomials} of $B$ if it is the set of the monomials
which are irreducible (non-divisible) by $B$
(see, e.g., \cite{CLO}, \cite{takayama1989}).
Let $g(z_1, \ldots, z_d)$ be a holonomic function
and we suppose that it is annihilated by a holonomic ideal $I$.
Let $S$ be the set of the standard monomials of a Gr\"obner basis of $RI$ in 
$R={\bf C}(z_1, \ldots, z_d)\langle\pd{1}, \ldots, \pd{d} \rangle$,
which is the ring of differential operators with rational function coefficients.
The cardinality of $S$ is finite and is called the {\it holonomic rank} of $I$.
We may suppose that $S$ contains $1$ as the first element of $S$.
Since the function $g$ is holonomic,
the column vector of functions $G=(s_i \bullet g \,|\, s_i \in S)^T$
satisfies the following set of linear partial differential equations
(see, e.g., \cite[p.39]{SST})
\begin{equation}  \label{eq:pfaffian}
 \frac{\partial G}{\partial {z_i}} = P_i G, \quad i=1, \dots, d
\end{equation}
where $P_i$ is a square matrix with entries in ${\bf C}(z_1, \ldots, z_d)$.
In fact, when the normal form of $\pd{i} s_m$ by $G$ in $R$ is
$\sum_n c^i_{mn} s_n$,
the rational functioin $c^i_{mn}$ is the $(m,n)$-th entry of the matrix
$P_i$ (see, e.g., the reductin algorithm in \cite{takayama1989}).
Note that each equation can be regarded as an ordinary differential equation
with respect to $z_i$ with parameters
$z_1, \ldots, z_{i-1}, z_{i+1}, \ldots, z_d$.
We call the system of differential equations (\ref{eq:pfaffian}) 
{\it the Pfaffian system (or equations)} for $g$.
The first entry of $G$, which is denoted by $G_1$, is $g$. 

A remarkable fact on holonomic function in this iteration scheme
is that
the gradient of $g$ and the Hessian of $g$ can be written in terms
of the vector function $G$,
which implies that we can evaluate the search direction for the gradient
descent from the value of $G$.
This fact is an easy consequence of the Gr\"obner basis theory,
but it is fundamental for the optimization of holonomic functions.
Precisely speaking, we have the following formula.
\begin{lemma}
\begin{enumerate}
\item Let $\sum_{s_j \in S} a_{ij} s_j$ be the normal form of 
$\pd{i} = \partial/\partial z_i$ 
by the Gr\"obner basis $B$ of $RI$ in $R$.
Here we have
$a_{ij} \in {\bf C}(z_1, \ldots, z_d)$. 
Let $A$ be the matrix with entries $a_{ij}$.
Then, we have 
$$  (\nabla g)(z^{(k)}) = A(z^{(k)}) G(z^{(k)}) $$
and
$$ (\nabla g)(z^{(k)}) = ( (P_1 G)_1, \ldots, (P_d G)_1 ) (z^{(k)}) $$
where $(v)_1$ notes the first entry of a vector $v$.
\item Let $\sum_k u_{ijk} s_k$ be the normal form of $\pd{i} \pd{j}$
with respect the Gr\"obner basis $B$
where
$u_{ijk} \in {\bf C}(z_1, \ldots, z_d)$. 
Then, we have
$$ \frac{\partial^2 g}{\partial z_i \partial z_j} (z^{(k)})
 = (u_{ij1}(z^{(k)}), \ldots, u_{ijd}(z^{(k)})) G(z^{(k)}) 
$$
and
$$ \frac{\partial^2 g}{\partial z_i \partial z_j} (z^{(k)})
 = \left( \left( \frac{\partial P_i}{\partial z_j} + P_i P_j \right) G \right)_1
$$
\end{enumerate}
\end{lemma}

{\it Proof}\/.
Since, $\pd{i} - \sum_j a_{ij} s_j \in RI$ and $(RI) \bullet g=0$,
we have
$ \partial_i \bullet g = \sum_{s_j \in S} a_{ij} (s_j \bullet g)$.
Then, we have the first identity of (1).
Since $\frac{\partial G}{\partial z_i} = P_i G$ and $G_1 = g$,
we have the second identity of (1).
The first identity of (2) can be shown analogously. 
Differentiating $\frac{\partial G}{\partial z_i} = P_i G$ by $z_j$,
we have
$ \frac{\partial^2 G}{\partial z_i \partial z_j}
 = \frac{\partial P_i}{\partial z_j} G + P_i \frac{\partial G}{\partial z_j}
 = \left( \frac{\partial P_i}{\partial z_j} G + P_i P_j \right) G
$.
Thus, the second identity of (2) is obtained.
\QED 

It follows from this lemma that we obtain the following
gradient descent for holonomic functions to find a local minimum.
We shortly call the method {\it holonomic gradient descent}\/.
Note that this is a symbolic-numeric algorithm.
\begin{algorithm} \rm   \label{algorithm:hgd0}
(Holonomic gradient descent)
\begin{enumerate}
\item Obtain a Gr\"obner basis of $RI$ in $R$ and the set of the standard monomials $S$ of the basis.
\item Compute the matrices $P_i$ in (\ref{eq:pfaffian}) by the normal form
algorithm and the Gr\"obner basis and the set of the standard monomials.
\item Compute the normal form $\partial_i$ by a Gr\"obner basis of $RI$
and determine the matrix $(a_{ij})$.
\item Take a point $z^{(0)}$ as a starting point 
and evaluate numerically the initial value of $G$ at $z=z^{(0)}$.
Denote the value by ${\bar G}$ and put $k=0$.
\item Evaluate numerically $(a_{ij}(z^{(k)})) {\bar G}$,
which is an approximate value of the gradient ${\tilde g} = \nabla g$ 
at $z^{(k)}$.
If a termination condition of the iteration is satisfied, then stop.
\item 
Put $z^{(k+1)} = z^{(k)} + h_k {\tilde g}$,
(move to $z^{(k)} + h_k {\tilde g}$). 
\item Obtain the approximate value of $G$ at $z=z^{(k+1)}$ 
by solving numerically the Pfaffian system (\ref{eq:pfaffian})
by the Runge-Kutta method
(see, e.g., \cite{OST}).
Set this value to ${\bar G}$. 
Increase the value of $k$ by $1$. Goto 5.
\end{enumerate}
Here, $h_k$ is the step length, which should be chosen by standard recipes
of gradient descent.
\end{algorithm}

Let us give two notes on numerical evaluations of $G$.
(1) The computation of the initial value $G$ requires a method 
depending on a given problem.
In case of the Fisher-Bingham integral, we use a numerical integration method.
(2) We use the Runge-Kutta method to evaluate $G$ at $z^{(k+1)}$ 
from the value of $G$ at $z^{(k)}$. 
Precisely speaking, we have
$$ \frac{d G(c(t))}{dt} 
=  \sum_{i=1}^d \frac{d c_i}{dt} \frac{\partial G}{\partial z_i}
=  \sum_{i=1}^d \left( \frac{d c_i}{dt} P_i\right) G
$$
for any smooth vector valued function $c(t)$.
We use this expression to numerically solve the Pfaffian system
to the direction $\tilde g$.

Elements of $P_i$ are rational functions.
The union of the zero sets of the denominators of elements of $P_i$'s
is called the {\it singular locus} of the Pfaffian equations (\ref{eq:pfaffian}). 
It is known that holonomic functions are holomorphic in the complement of the singular locus
of corresponding Pfaffian equations.
We can apply known convergence criteria to this algorithm (see, e.g., \cite{synman}) when we look for a local minimum in a connected domain 
in the complement of the singular locus.
Hence, we have to limit the search domain of a local minimum in the connected domain.

The holonomic gradient descent can be applied to a large class
of optimization problems.
It is well known that when $f$ and $g$ are holonomic functions,
then the sum $f+g$ and the product $fg$ are also holonomic functions.
A remarkable fact is that when $f$ is a holonomic function 
in $z_1, \ldots, z_d$, then the definite integral
$ \int_{a_d}^{b_d} f(z_1, \ldots, z_d) dz_d $
is also a holonomic function in $z_1, \ldots, z_{d-1}$.
We have algorithms to find systems of differential equations
for the sum, the product, and the definite integral.
As to these topics, see, e.g., \cite{APZ}, \cite{NN}, \cite{oaku}, \cite{OST}, \cite{SST} and their references.
It follows from these results that
we can present our algorithm in the following form.

\begin{algorithm} \rm \label{algorithm:hgdi} \quad 
(Holonomic gradient descent for integrals)\\
Input: a definite integral $F(z) = \int_C f(z,t) dt$ with parameters $z=(z_1, \ldots, z_d)$
where $f(z,t)$ is a holonomic function of which annihilating ideal is $J$. \\
\quad\quad A holonomic function $g(z)$ of which annihilating ideal is $J'$. \\
Output: An approximate local minimum of $g(z) F(z)$ for $z \in E$.
\begin{enumerate}
\item Apply integration algorithms for the holonomic ideal $J$ 
(see, e.g., \cite{APZ}, \cite{NN}, \cite{oaku}, \cite{OST}, \cite{SST} and their references)
to find a holonomic ideal $\int J$ annihilating the function $F(z)$.
\item Obtain a holonomic ideal $I$ which annihilates $g(z) F(z)$ from $\int J$ and $J'$
(see, e.g., \cite{takayama1992}, \cite{OST}).
\item Apply Algorithm \ref{algorithm:hgd0} for $I$
where starting values of $F(z)$ and its derivatives are computed by a numerical integration method.
\end{enumerate}
We note that integration algorithms require some conditions for the domain 
of the integration $C$.
The domain $C$ must satisfy  the conditions.
For example, when $C$ is a product of segments and $C$ is contained
in the complement of the singularities of $f(z,t)$,
the domain satisfies the conditions.
The search domain $E$ must be in the complement of the singular locus 
of the Pfaffian equations
for $g(z) F(z)$.
\end{algorithm} 

Let us illustrate our method with a small sized problem.
\begin{example} \rm \label{example:airy}
$d=1$, $z=x$.
$g(x) = \exp(-x+1) \int_0^\infty \exp(x t - t^3) dt$.
The function $g(x)$ satisfies the differential equation
$ (3 \partial_x^2 + 6 \partial_x + (3-x))\bullet  g = \exp(-x+1)$,
which can be obtained by an integration algorithm for $D$-modules
\cite{NN}.
The holonomic rank is $2$ and
we use a set of standard monomials $S=\{1, \partial_x\}$
and we have
$$ \frac{d G}{dx}
     = \pmatrix{ 0   & 1 \cr
                (-3+x)/3 & -2 \cr} G + \pmatrix{ 0 \cr \exp(-x+1)/3 \cr}
$$ 
This system is obtained by the normal form algorithm
in the ring $R$ \cite{yang}.
We note that it is easy to generalize
our algorithm for a holonomic function which satisfies
inhomogeneous holonomic system. 
Note that $\frac{dg}{dx}=\nabla g = \pmatrix{0 & 1} G$.
We evaluate 
$G(0)=(g(0),g'(0))^T$  by a numerical integration method;
$\bar G(0) = (2.427, -1.20)^T$.
We apply the holonomic gradient descent in the search domain $E=[0,5]$ with 
$h_k=-0.1, H_k = 1$
and the 4th order Runge-Kutta method
and obtain $x=e=3.4$ and $g(e)=1.016$
as the minimum in this domain.
\end{example}
The holonomic gradient descent is nothing but Euler's method 
when the number of variables is $1$.

As we have seen,
by utilizing integration algorithms,
we can apply the holonomic gradient descent for a large class of
optimization problems including integrals with parameters.
However, integration algorithms require huge computational resources
and we can solve only relatively small sized problems.
Therefore, if we want to apply our method to larger problems
for holonomic functions,
we need to find systems of differential equations and
Pfaffian equations without utilizing general algorithms.
In fact, we will study a system of differential equations and Pfaffian equations
for the Fisher-Bingham integral in the following sections
to apply our method to a maximal likelihood estimate problem.

\section{Fisher-Bingham Integral on $S^n$}  \label{sec:fbi}

We denote by $S^n(r)$ the $n$-dimensional sphere with the radius $r$
in the $n+1$ dimensional Euclidean space.
Let $x$ be a $(n+1) \times (n+1)$ symmetric matrix
and $y$ a row vector of length $n+1$.
We are interested in the following integral with the parameters $x,y,r$.
\begin{equation} \label{eq:fbi}
  F(x,y,r) = \int_{S^n(r)} \exp( t^T x t + y t ) |dt|
\end{equation}
Here, $t$ is the column vector $(t_1, \ldots, t_{n+1})^T$
and $|dt|$ is the standard measure on the sphere.
For example,
in case of $n=1$, the measure $|dt|$ is $r d\theta$ in the polar coordinate system
$t_1 = r \cos \theta, t_2 = r \sin \theta$.
We call the integral (\ref{eq:fbi})
{\it the Fisher-Bingham integral} on the sphere $S^n(r)$.

We denote by $x_{ii}$  the $i$-th diagonal entry of the matrix $x$
and by $x_{ij}/2$ the $(i,j)$-th entry (or $(j,i)$-th entry) of the matrix $x$.
Then, we can regard the function (the Fisher-Bingham integral) 
$F(x,y,r)$ as the function
of $x_{ij}$  ($1 \leq i \leq j \leq n+1$) and 
$y_i$ ($1 \leq i \leq n+1$) and $r$.

\begin{theorem}
The Fisher-Bingham integral $F(x,y,r)$ is a holonomic function.
\end{theorem}

{\it Proof}\/. 
We will prove it for $n=1$ to avoid complicated indices.
The cases for $n > 1$ can be shown analogously.

Put $x_1=r \cos \theta, x_2=r \sin \theta$ (the polar coordinate system).
Then, the invariant measure $|dt|$ is written as $r d\theta$.
Therefore, 
$F(x,y,r)= \int_0^{2 \pi} e^{g(x,y,r,\theta)} r d\theta$
where 
$g(x,y,r,\theta) = x_{11} r^2 \cos^2 \theta + x_{12} r^2 \cos \theta \sin \theta 
                 + x_{22} r^2 \sin^2 \theta + y_1 r \cos \theta + y_2 r \sin \theta$.
If we put $s=\tan\frac{\theta}{2}$,
then $\sin \theta=2s/(s^2+1)$ and $\cos \theta = (1-s^2)/(s^2+1)$
and $d\theta = \frac{2}{1+s^2} ds$
(rational representation of trigonometric functions).
Then, the integral $F(x,y,r)$ can be written as
$$\int_{-\infty}^\infty h(x,y,r,s)  ds , \quad
h=e^{{\tilde g}(x,y,r,s)} \frac{2}{1+s^2}
$$
where ${\tilde g}$ is a rational function in $x,y,r,s$.
It is known that the exponential of a rational function is a holonomic function
and the product of holonomic functions is a holonomic function,
then the integrand is a holonomic function in $x,y,r,s$ (see, e.g., \cite{OST} and \cite{OTW}).
By Lemma \ref{ap:lemma:holonomicElimination} in the Appendix,
there exists a differential operator 
$ \ell(x,y,r, \partial_{x_{ij}}) - \partial_s \ell_1(x,y,r, \partial_{x_{ij}},\partial_s)$
depending only on
$x, \partial_{x_{ij}}, y, r, \partial_s$
which annihilates the integrand $h$.
Therefore, we have $\ell \bullet F(x,y,r) = [ \ell_1 \bullet h]_{-\infty}^\infty$. 
Since we can show that $\partial_{x_{ij}}^m \partial_s^n \bullet h $ 
is a finite holonomic function at $s=\pm \infty$
for any non-negative integers $m$ and $n$,
the function $F(x,y,r)$ is annihilated by an ordinary differential operator of $\partial_{x_{ij}}$
with parameters $x,y,r$.
The existence of annihilating ordinary differential operators with respect to
$\partial_{y_i}$ and $\partial_r$ can be shown analogously.
This existence implies that $F(x,y,r)$ is a holonomic function 
(see, e.g., \cite[Theorem 2.4]{takayama1992}).
\QED

\section{Holonomic system for the Fisher-Bingham Integral} \label{sec:hsfbi}

In Example \ref{example:airy}, we obtained a differential equation 
for the definite integral with parameters by a D-module algorithm.
This algorithm works for any definite integral with a holonomic integrand,
however, it requires huge computational resources.
For the Fisher-Bingham integral, we can obtain a holonomic system of
differential equations for the case of $n=1$ by our computer program.
The case of $n=2$ is not feasible by our program.
We obtain the following result for general $n$ by utilizing
an invariance of the Fisher-Bingham integral.

\begin{theorem} \label{theorem:eqs}
The function $ F(x,y,r)$ is annihilated by the following system of 
linear partial differential operators.
\begin{eqnarray}
&& \partial_{x_{ij}} - \partial_{y_i} \partial_{y_j}, \quad (i \leq j) \label{eq:toric}\\
&& \sum_{i=1}^{n+1} \partial_{x_{ii}} - r^2, \\
&& x_{ij} \partial_{x_{ii}} + 2 (x_{jj}-x_{ii}) \partial_{x_{ij}}
  - x_{ij} \partial_{x_{jj}} 
  + \sum_{k \not= i,j} (x_{jk} \partial_{x_{ik}} - x_{ik} \partial_{x_{jk}}) \nonumber \\
&&\quad   + y_j \partial_{y_i} - y_i \partial_{y_j}, \quad (i<j, x_{k\ell}=x_{\ell k}), \label{eq:orthogonal}\\
&& r \partial_r - 2 \sum_{i \leq j} x_{ij} \partial_{x_{ij}} - \sum_i y_i \partial_{y_i} - n. \label{eq:scale}
\end{eqnarray}
\end{theorem}
We note that operators of the form (\ref{eq:toric}) can be written as
$$ \partial^u - \partial^v, \quad Au=Av,\  u,v \in {\bf N}^{(n+1)(n/2+2)}. $$
Here, $A$ is the support matrix of the polynomial 
$ t^T x t + y t $ with respect to $t$.
For example, in case of $n=1$, the polynomial is
$ x_{11} t_1^2 + x_{12} t_1 t_2 + x_{22} t_2^2 + y_1 t_1 + y_2 t_2$
and the matrix $A$ is
$$ A=\pmatrix{
    2 & 1 & 0 & 1 & 0 \cr
    0 & 1 & 2 & 0 & 1 \cr
  }
$$
of which column vectors stand for supports of the polynomial respectively.

{\it Proof}\/.
Denote by $g(x,y,t)=\exp(t^Txt+yt)$ the integrand of (\ref{eq:fbi}).
The operator $\partial_{x_{ij}}-\partial_{y_i}\partial_{y_j}$ annihilates $g(x,y,t)$
because $(\partial_{x_{ij}}-\partial_{y_i}\partial_{y_j})\bullet g
=(t_it_j-t_it_j)g=0$.
On the sphere $S^n(r)$, we have an identity $\sum_{i=1}^{n+1}t_i^2=r^2$.
Hence $\sum_{i=1}^{n+1}\partial_{x_{ii}}-r^2$ annihilates $g(x,y,t)$
for $t\in S^n(r)$.

Let us prove  (\ref{eq:orthogonal}).
By the invariance of the measure $|dt|$ with respect to the orthogonal
group,
we have $F(PxP^T,yP^T,r)=F(x,y,r)$ for any orthogonal 
transformation $P$ on $S^n(r)$.
Let $I_{n+1}$ be the $(n+1)\times (n+1)$ identity matrix
and $e_{ij}$ be an $(n+1)\times (n+1)$ matrix whose $(k,l)$-th entry $(e_{ij})_{kl}$
is $1$ if $(i,j)=(k,l)$ and $0$ else.
Put $P=\pmatrix{ \cos \epsilon & -\sin \epsilon \cr
                 \sin \epsilon & \cos \epsilon \cr} \oplus I_{n-1}$.
This is an $(n+1) \times (n+1)$ orthogonal matrix and we have
$P=I_{n+1}+\epsilon (e_{12}-e_{21})+O(\epsilon^2)$.
Hence we have
\begin{eqnarray*}
 PxP^T
 &=&
 (I+\epsilon(e_{12}-e_{21}))x(I+\epsilon(e_{21}-e_{12})) + O(\epsilon^2)
 \\
 &=& x + \epsilon (e_{12}x-e_{21}x+xe_{21}-xe_{12}) + O(\epsilon^2)
  \\
 &=& x + \epsilon \sum_{i\le j}f_{ij}(x)(e_{ij}+e_{ji})/2+O(\epsilon^2),
\end{eqnarray*}
where
\[
 f_{ij}(x)
 = \left\{
 \begin{array}{lll}
  x_{12}& {\rm if}& i=j=1,
   \\
  2(x_{22}-x_{11})& {\rm if}& i=1,j=2,
   \\
  -x_{12} & {\rm if}& i=j=2,
   \\
  x_{2j}& {\rm if}& i=1,j\ge 3,
   \\
  -x_{1j}& {\rm if}& i=2,j\ge 3,
   \\
  0& {\rm if}& j\ge i\ge 3,
 \end{array}
 \right.
\]
and
\[
 yP^T = 
 y + \epsilon
 \pmatrix{
 y_2& -y_1& 0
 }
 +O(\epsilon^2).
\]
Differentiating the identity $F(PxP^T,yP^T,r)-F(x,y,r)=0$ by $\epsilon$,
we obtain
\begin{eqnarray*}
 0
 &=& \left(\sum_{i\le j}f_{ij}(x)\partial_{x_{ij}}
   +y_2\partial_{y_1}-y_1\partial_{y_2}\right)\bullet F + O(\epsilon).
\end{eqnarray*}
Taking the limit $\epsilon \rightarrow 0$, we have 
(\ref{eq:orthogonal}) with $i=1$ and $j=2$.
By symmetry we have (\ref{eq:orthogonal}) for any $i<j$.

Finally we differentiate the identity 
$\rho^n F(\rho^2 x,\rho y,r)=F(x,y,\rho r)$ by $\rho$
and take the limit $\rho \rightarrow 1$.
Then, we obtain
\begin{eqnarray*}
 \left(n + 2\sum_{i\le j}x_{ij}\partial_{x_{ij}} + \sum_{i}y_i\partial_{y_i}
 \right)\bullet F
 &=& r\partial_r\bullet F
\end{eqnarray*}
This shows that $F$ is annihilated by (\ref{eq:scale}).
\QED

\begin{example} \rm
When $n=1$, the system is written as follows.
\begin{eqnarray*}
  && \partial_{x_{11}}-\partial_{y_1}^2, 
  \partial_{x_{12}}-\partial_{y_1}\partial_{y_2},
  \partial_{x_{22}}-\partial_{y_2}^2,
  \\
  && \partial_{x_{11}}+\partial_{x_{22}}-r^2,
  \\
  && x_{12}\partial_{x_{11}}+2(x_{22}-x_{11})\partial_{x_{12}} - x_{12}\partial_{x_{22}} + y_2\partial_{y_1} - y_1\partial_{y_2},
  \\
  && r\partial_r-2(x_{11}\partial_{x_{11}}+x_{12}\partial_{x_{12}}+x_{22}\partial_{x_{22}})-(y_1\partial_{y_1}+y_2\partial_{y_2})-1.
\end{eqnarray*}
\end{example}

\begin{example} \rm
When $n=2$, the system is written as follows.
\begin{eqnarray*}
&&  \partial_{{x}_{11}}-  \partial_{{y}_{1}}^{ 2} ,  \partial_{{x}_{12}}-  \partial_{{y}_{1}}  \partial_{{y}_{2}},  \partial_{{x}_{13}}-  \partial_{{y}_{1}}  \partial_{{y}_{3}},  \\
&& \partial_{{x}_{22}}-  \partial_{{y}_{2}}^{ 2} ,  \partial_{{x}_{23}}-  \partial_{{y}_{2}}  \partial_{{y}_{3}},  \partial_{{x}_{33}}-  \partial_{{y}_{3}}^{ 2} , \\
&&   \partial_{{x}_{11}}+ \partial_{{x}_{22}}+ \partial_{{x}_{33}}-  {r}^{ 2} ,  \\
&&      {x}_{12}  \partial_{{x}_{11}}+   2  (  {x}_{22}- {x}_{11})  \partial_{{x}_{12}}-  {x}_{12}  \partial_{{x}_{22}}+  {x}_{23}  \partial_{{x}_{13}}-  {x}_{13}  \partial_{{x}_{23}}+  {y}_{2}  \partial_{{y}_{1}}-  {y}_{1}  \partial_{{y}_{2}}, \\
&&       {x}_{13}  \partial_{{x}_{11}}+   2  (  {x}_{33}- {x}_{11})  \partial_{{x}_{13}}-  {x}_{13}  \partial_{{x}_{33}}+  {x}_{23}  \partial_{{x}_{12}}-  {x}_{12}  \partial_{{x}_{23}}+  {y}_{3}  \partial_{{y}_{1}}-  {y}_{1}  \partial_{{y}_{3}},  \\
&&      {x}_{23}  \partial_{{x}_{22}}+   2  (  {x}_{33}- {x}_{22})  \partial_{{x}_{23}}-  {x}_{23}  \partial_{{x}_{33}}+  {x}_{13}  \partial_{{x}_{12}}-  {x}_{12}  \partial_{{x}_{13}}+  {y}_{3}  \partial_{{y}_{2}}-  {y}_{2}  \partial_{{y}_{3}},  \\
&&  {r}  \partial_{{r}}-  2  (       {x}_{11}  \partial_{{x}_{11}}+  {x}_{12}  \partial_{{x}_{12}}+  {x}_{13}  \partial_{{x}_{13}}+  {x}_{22}  \partial_{{x}_{22}}+  {x}_{23}  \partial_{{x}_{23}}+  {x}_{33}  \partial_{{x}_{33}})\\
&& \quad- (    {y}_{1}  \partial_{{y}_{1}}+  {y}_{2}  \partial_{{y}_{2}}+  {y}_{3}  \partial_{{y}_{3}})-2.
\end{eqnarray*}
\end{example}

\begin{proposition}
\begin{enumerate}
\item The operators given in Theorem \ref{theorem:eqs} generate
a holonomic ideal in case of $n=1$ and $n=2$.
\item The holonomic rank of the system for $n=1$ is $4$.
A set of standard monomials in $R$  is
$$ 1,  \partial_{y_1} ,  \partial_{y_2},  \partial_r.$$
\item The holonomic rank of the system for $n=2$ is $6$.
A set of standard monomials in $R$  is
   $$ 1,   \partial_r,  \partial_{y_3},  \partial_{y_2}, 
    \partial_{y_1} ,  \partial_{x_{33}}.  $$
\end{enumerate}
\end{proposition}

The proposition can be shown by a calculation on a computer
with applying algorithms for holonomic systems 
\cite{asir}, \cite[{\tt toc.html}]{web}, \cite{SST}.

We conjecture that the system of operators given in Theorem \ref{theorem:eqs}
generates a holonomic ideal in $D$,
which is the ring of differential operators with polynomial coefficients.
We can prove weaker result that they generate a zero dimensional ideal
in $R$, which is sufficient for applying the holonomic graident.
This result can also be used to derive Pfaffian equations.
We will prove the zero dimensionality in the sequel.

For the Fisher-Bingham integral $F(x,y,r)$,
let $X=\{x,y,r\}$ be the set of all variables and $\partial_X$
be the corresponding differential operators.
Consider a ring $R={\bf C}(X)\langle \partial_X\rangle$.
Let $I\subset R$ be the ideal generated by the operators
(\ref{eq:toric}) -- (\ref{eq:scale}) annihilating $F(x,y,r)$ (Theorem~\ref{theorem:eqs}).
We show that the ideal $I$ is zero-dimensional, that is,
the quotient space $R/I$ is a finite-dimensional vector space over ${\bf C}(X)$.

We denote $\partial_{ij}=\partial_{x_{ij}}$ and $\partial_{i}=\partial_{y_{i}}$ for simplicity.
The symbol $\partial_r$ is reserved for $\partial/\partial r$.
It is easy to see that $I$ is generated by
\begin{eqnarray}
 A_{ij} &=& \partial_{ij} - \partial_{i}\partial_{j},
 \label{eq:proof-zero-dim-1}\\
 B &=& \sum_i\partial_i^2 - r^2,
 \label{eq:proof-zero-dim-2}\\
 C_{ij} &=& 2(x_{jj}-x_{ii})\partial_i\partial_j + x_{ij} \partial_i^2 - x_{ij}\partial_j^2
  \nonumber\\
 && \quad + \sum_{k\neq i,j}(x_{jk}\partial_i\partial_k - x_{ik}\partial_j\partial_k)
 +y_j\partial_i-y_i\partial_j,
 \label{eq:proof-zero-dim-3}\\
 E &=& r\partial_r-2\sum_{i\leq j}x_{ij}\partial_i\partial_j - \sum_iy_i\partial_i - n.
 \label{eq:proof-zero-dim-4}
\end{eqnarray}
We write $\ell_1\equiv \ell_2$ if $\ell_1-\ell_2\in I$.

\begin{theorem} \label{theorem:zero-dim}
  Put $S=\{1,\partial_1,\ldots,\partial_{n+1},\partial_1^2,\ldots,\partial_n^2\}$ and
  let $L_S$ be the vector space over ${\bf C}(X)$ spanned by $S$.
  Then we have $R=L_S+I$.
  In particular, the ideal $I$ is zero-dimensional.
\end{theorem}

We prepare two lemmas.
The proof is given later.

\begin{lemma} \label{lemma:proof-zero-dim-1}
 For any $i$ and $j$, we have $\partial_i\partial_j\in L_S+I$.
\end{lemma}

\begin{lemma} \label{lemma:proof-zero-dim-2}
 For any $i$, $j$ and $k$, we have $\partial_i\partial_j\partial_k\in L_S+I$.
\end{lemma}

We give a proof of Theorem~\ref{theorem:zero-dim} by using the lemmas.
The proof implicitly uses a lexicographic order $\prec$
such that $\partial_k\prec\partial_{ij}$
and $\partial_k\prec\partial_r$ for any $k,i,j$.
\bigbreak

{\it Proof of Theorem~\ref{theorem:zero-dim}}\/.
We first show that $R={\bf C}(X)\langle\partial_1,\ldots,\partial_{n+1}\rangle+I$.
Let $f$ be an element of $R$.
If a term of $f$ is written as $g\partial_{ij}$ with $g\in R$,
then we can replace $g\partial_{ij}$ with $g\partial_i\partial_j$
because $\partial_{ij}\equiv \partial_i\partial_j$.
By induction, there exists some $f'\in R$ without $\partial_{ij}$
such that $f\equiv f'$.
If $f'$ contains $\partial_r$,
we can replace $\partial_r$ with a polynomial of $\{\partial_k\}$
by the annihilator (\ref{eq:proof-zero-dim-4}).
By induction, there exists some $f''\in {\bf C}(X)\langle\partial_1,\ldots,\partial_{n+1}\rangle$
such that $f\equiv f'\equiv f''$.
This proves $R={\bf C}(X)\langle \partial_1,\ldots,\partial_{n+1}\rangle+I$.
Now we show that ${\bf C}(X)\langle \partial_1,\ldots,\partial_{n+1}\rangle+I=L_S+I$.
Let $f=\prod_{i=1}^{n+1}\partial_i^{\beta_i}$ be any monomial in ${\bf C}(X)\langle \partial_1,\ldots,\partial_{n+1}\rangle$
with the total degree $|\beta|=\sum_{i=1}^{n+1}\beta_i$.
If $|\beta|\leq 1$, clearly $f\in L_S\subset L_S+I$.
If $|\beta|=2$, Lemma~\ref{lemma:proof-zero-dim-1} shows $f\in L_S+I$.
If $|\beta|\geq 3$, then by Lemma~\ref{lemma:proof-zero-dim-2}
there is $f'$ with the total degree less than or equal to $|\beta|-1$
such that $f\equiv f'$.
By induction, we have some $f'$ with the total degree less than or equal 
to $2$ such that $f\equiv f'$ ($\in L_S+I$).
This proves Theorem~\ref{theorem:zero-dim}.
Q.E.D.
\bigbreak

Now we prove Lemma~\ref{lemma:proof-zero-dim-1} and Lemma~\ref{lemma:proof-zero-dim-2}.

\bigbreak
{\it Proof of Lemma~\ref{lemma:proof-zero-dim-1}}\/.
  From the definition of $S$,
  it is obvious that $\partial_i^2\in L_S$ for $1\leq i\leq n$.
  Since $\partial_{n+1}^2\equiv -\sum_{i=1}^n\partial_i^2+r$ by (\ref{eq:proof-zero-dim-2}),
  we have $\partial_{n+1}^2\in L_S+I$.
  Now we prove that $\partial_i\partial_j\in L_S+I$ for any $1\leq i<j\leq n+1$.
  We use the annihilator $C_{ij}$ in (\ref{eq:proof-zero-dim-3}).
 Denote the quadratic part of $C_{ij}$ by $\sum_{k<l}P_{ij,kl}\partial_k\partial_l$,
 where $P_{ij,kl}=P_{ij,kl}(x,y,r)\in {\bf C}(X)$.
 Since $1$ and $\partial_k$ are in $L_S+I$, we have
 \[
    \sum_{k<l} P_{ij,kl}\partial_k\partial_l
    \in L_S+I.
 \]
 To show $\partial_i\partial_j\in L_S+I$,
  it is sufficient to prove that the determinant of the coefficient matrix $(P_{ij,kl})_{i<j;k<l}$
  is a non-zero element in ${\bf C}(X)$.
  We evaluate $P_{ij,kl}$ at a point $(x,y,r)=(\bar{x},\bar{y},\bar{r})$ such that
  $\bar{x}_{ii}\neq \bar{x}_{jj}$ and $\bar{x}_{ij}=0$ for any $i<j$. Then we obtain
 \[
  P_{ij,kl}(\bar{x},\bar{y},\bar{r})
  = \left\{
  \begin{array}{ll}
   2(\bar{x}_{jj}-\bar{x}_{ii})& \mbox{if}\ (i,j)=(k,l), \\
   0& \mbox{else}.
  \end{array}
  \right.
 \]
 In particular, $P_{ij,kl}(\bar{x},\bar{y},\bar{r})$ is a diagonal matrix
 and its determinant is $\prod_{i<j}2(\bar{x}_{jj}-\bar{x}_{ii})\neq 0$.
 Hence the determinant of $(P_{ij,kl})$ is non-zero in ${\bf C}(X)$.
Q.E.D.
\bigbreak

{\it Proof of Lemma~\ref{lemma:proof-zero-dim-2}}\/.
  Consider an operator $\partial_i\partial_j\partial_k$ with $i\leq j\leq k$.
  If $j=k=n+1$, then $\partial_i\partial_{n+1}^2\equiv
  \partial_i(-\sum_{l=1}^n\partial_l^2 + r^2)$.
  Hence we can assume $j\leq n$.
  By using the operator $C_{ij}$ in (\ref{eq:proof-zero-dim-3}),
 we define an operator $G_{ijk}$ by
  \[
    G_{ijk} = 
    \left\{
     \begin{array}{ll}
       \partial_iC_{jk} & \mbox{if}\ j<k,  \\
       \partial_jC_{ij} & \mbox{if}\ i<j=k (\leq n),\\
       \partial_{n+1}C_{i,n+1} & \mbox{if}\ i=j=k (\leq n)
     \end{array}
    \right.
  \]
 Then $G_{ijk}\equiv 0$.
 As in the proof of Lemma~\ref{lemma:proof-zero-dim-1},
 denote the cubic term of $G_{ijk}$ by
 $\sum_{a\leq b\leq c;b\leq n}P_{ijk,abc}\partial_a\partial_b\partial_c$.
 Since all quadratic terms are in $L_S+I$, we obtain
 \[
  \sum_{a\leq b\leq c;b\leq n}P_{ijk,abc}\partial_a\partial_b\partial_c
 \in L_S+I.
 \]
 It is sufficient to show that $\det(P_{ijk,abc})$ is a non-zero element 
 in ${\bf C}(X)$.
 As in the proof of Lemma~\ref{lemma:proof-zero-dim-1},
 we evaluate $P_{ijk,abc}$ at a point $(\bar{x},\bar{y},\bar{r})$
 such that $\bar{x}_{ii}\neq \bar{x}_{jj}$ and $\bar{x}_{ij}=0$
 for any $i<j$. Then, with a little effort, we obtain
 \begin{eqnarray*}
  \lefteqn{P_{ijk,abc}(\bar{x},\bar{y},\bar{r})} \\
  &=& \left\{
  \begin{array}{ll}
   2(\bar{x}_{kk}-\bar{x}_{jj})\delta_{ia}\delta_{jb}\delta_{kc}&
    \mbox{if}\ j<k,\\
   2(\bar{x}_{jj}-\bar{x}_{ii})\delta_{ia}\delta_{jb}\delta_{jc}&
    \mbox{if}\ i<j=k(\leq n),\\
   -2(\bar{x}_{n+1,n+1}-\bar{x}_{ii})\{\delta_{ia}\delta_{ib}\delta_{ic} & \\
   \quad +\sum_{h<i}\delta_{ha}\delta_{hb}\delta_{ic} + \sum_{i<h\leq n}\delta_{ia}\delta_{hb}\delta_{hc}
  \}& 
  \mbox{if}\ i=j=k(\leq n).
  \end{array}
  \right.
 \end{eqnarray*}
 Remark that all the diagonal elements $P_{ijk,ijk}$ are non-zero.
 We sort indices $\{(i,j,k)\mid i\leq j\leq k,j\leq n\}$ in such a way that
 $(i,i,i)$ is greater than $(j,k,l)$ unless $j=k=l$.
 Then we can conclude that $P_{ijk,abc}(\bar{x},\bar{y},\bar{r})=0$ if $(i,j,k)$ is less than $(a,b,c)$.
 Hence $P_{ijk,abc}(\bar{x},\bar{y},\bar{r})$ is a triangular matrix
 and its determinant is product of the diagonal elements.
 This proves that $\det(P_{ijk,abc})$ is a non-zero element in ${\bf C}(X)$.
\QED

\section{Computational Results} \label{sec:cr}

Let us apply the holonomic gradient descent to minimize the holonomic function
\begin{eqnarray}
  F(x,y,1)\exp\left(-\sum_{1\le i\le j\le n}S_{ij}x_{ij}-\sum_{i}S_iy_i\right)
  \label{eqn:obj}
\end{eqnarray}
with respect to $x$ and $y$ for given data $((S_{ij})_{i\le j},(S_i))$.
Here $F(x,y,1)$ is the Fisher-Bingham integral (\ref{eq:fbi})
with $r=1$.

First we describe the background in statistics.
This paragraph can be skipped for the reader interested only in computational results.
{\it The Fisher-Bingham family} on the sphere $S^n(1)$ is defined by
the set of probability density functions
\begin{equation} \label{eq:fb-dens}
 p(t|x,y)
  = F(x,y,1)^{-1}\exp(t^{\top}xt+yt)
\end{equation}
with respect to the standard measure $|dt|$ on $S^n(1)$.
Since $\int_{S^n(1)}p(t|x,y)|dt|=1$, 
the function $p(t|x,y)$ is actually a probability density function.
We note that the parameter $x$ has redundancy.
In fact, for any real number $c$ the density function
$p(t|x+cI,y)$ is equal to $p(t|x,y)$, where $I$ denotes the identity matrix.
{\it A sample} refers to a set of points $\{t(1),\ldots,t(N)\}$
on $S^n(1)$, where $N\ge 1$ is called the sample size.
Assume that the sample is distributed according to
$\prod_{\nu=1}^Np(t(\nu)|x,y)$ (independently identically distributed).
To estimate the unknown parameter $(x,y)$
from the sample is a main problem in statistics.
An established method is {\it the maximum likelihood method (MLE)}
that maximizes a function $\prod_{\nu=1}^N p(t(\nu)|x,y)$ with respect to $(x,y)$.
The MLE is equivalent to minimize the function (\ref{eqn:obj})
with $S_{ij}=N^{-1}\sum_{\nu=1}^Nt_i(\nu)t_j(\nu)$ and $S_i=N^{-1}\sum_{\nu=1}^Nt_i(\nu)$.
It is known that the logarithm of (\ref{eqn:obj}) is convex (see e.g.\ \cite{Ba})
and therefore a local minimum at an interior point is actually the global minimum.
Although gradient systems on probability families for optimization are considered by \cite{nakamura},
difficulty of computing the integral $F$ is not taken into account.
See \cite{MJ} for details on the Fisher-Bingham family and other probability families on the sphere.
We test two examples, astronomical data and magnetism data.
The astronomical data consist of the locations of 188 stars of magnitude
brighter than or equal to 3.0.
The data is available from
the Bright Star Catalog (5th Revised Ed.)
distributed from the Astronomical Data Center.
The magnetism data is analyzed in \cite{creer59} and \cite{kent82}.

The data and programs to test the following examples
can be obtained from \cite{web}. 

\begin{remark} \rm
Let $e_i$ be the $i$-th standard vector.
We note that
$G(z^{(k)}+e_i h_k)$ can approximately be obtained by
evaluating $P_i(z^{(k)}) G(z^{(k)}) h_k$.
In our implementation in \cite{web},
we choose a search direction $d^{(k)}$ which is parallel to a coordinate axis.
In other words, 
if the direction $h_k e_i$ is chosen, then  we move to the direction
as long as $g$ decreases to the direction $h_k e_i$.
Because  $P_i$ is a matrix of a huge size
and the computational cost of restricting the variables $z_j$, $j \not= i$
in $P_i$ to numbers is extremely high
in the problem of Fisher-Bigham integral and our implementation.
\end{remark}

\begin{figure}[tbh]
\includegraphics[width=12cm]{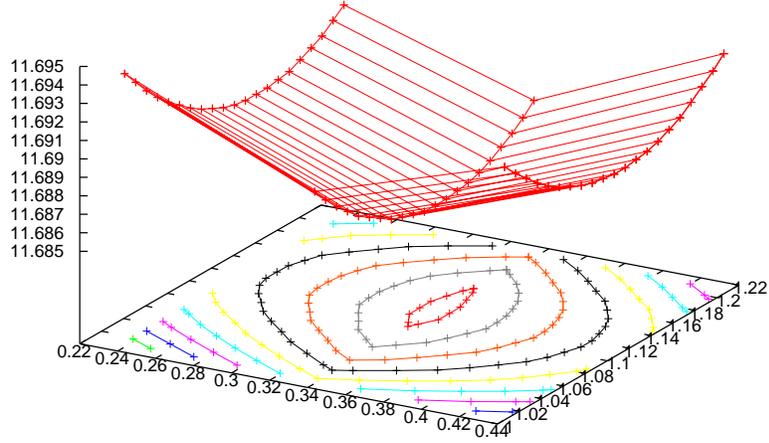}
\caption{Graph of the target function with varying $x_{12}$ and $x_{13}$ around the minimal point for astronomical data.}
\label{fig:astro}
\end{figure}

\bigbreak
\noindent
{\bf Astronomical data}:
We consider the problem to 
minimize 
$$
F(x,y,1)   \exp\left( - \sum_{1 \leq i \leq  j \leq 3} S_{ij} x_{ij} 
                      - \sum_{i} S_i y_i \right)
$$
on 
\begin{eqnarray*}
& & (x_{11}, x_{12}, x_{13}, x_{22}, x_{23}, x_{33}, y_1,y_2, y_3) \\
&\in &E=[-30,10] \times  [-30,10] \times [-30,10]
                \times [-30,10] \times [-30,20]
                                 \times [-30,-0.01] \\
 & &  \quad\quad
      \times [-30,-0.01] \times [-30,-0.001] \times [-30,10]
\end{eqnarray*}
where 
\begin{eqnarray*}
&&
  (S_{11}, S_{12}, S_{13}, 
           S_{22}, S_{23},
                   S_{33},
   S_{1}, S_{2}, S_{3}) \\
&=&
 (0.3119,0.0292,0.0707,
         0.3605,0.0462,
                 0.3276,
   -0.0063,-0.0054,-0.0762).
\end{eqnarray*}
The result is that
the minimum $11.68573121328159669$ is taken at \\ 
$x=\pmatrix{ -0.161 & 0.3377/2 & 1.1104/2 \cr
            0.3377/2 & 0.2538  & 0.6424/2 \cr
            1.1104/2 & 0.6424/2  & -0.0928 \cr }$,
$y=(\underline{-0.019}, \underline{-0.0162},-0.2286)
$
with the grid size $0.05$ and the 4th order Runge-Kutta method 
for solving the Pfaffian system numerically
(see Fig.~\ref{fig:astro}), where the values near the border are underlined.
A starting point is found by a quadratic approximation of $F(x,y,1)$,
which is exactly calculated from the moments
of the uniform distribution on the sphere,
and solving the optimization problem for the quadratic polynomial.

We briefly discuss the statistical meaning of the result.
The spectral decomposition of $x$ is
$x = \sum_{i=1}^3\lambda_iz_iz_i^T$ with
\[
 (\lambda_1,\lambda_2,\lambda_3) = (0.7047,-0.0103,-0.6944)
\]
and
\[
 (z_1,z_2,z_3) = 
 \pmatrix{
 -0.5063&  0.5055&  0.6987\cr
 -0.6181& -0.7777&  0.1148\cr
 -0.6014&  0.3737& -0.7061
 }.
\]
From the decomposition the density function (\ref{eq:fb-dens})
is high around $\pm z_1$ and low around $\pm z_3$.
The effect of $y$ is small because $|y|=0.230$ is smaller than $|\lambda_i|$'s.

As we have seen, we have determined the model parameters $x$ and $y$
by the holonomic graident descent successfully.
However, the computation poses us two future problems to make
the method stronger and more useful.
The first problem is to determine the search domain $E$
of $x$ and $y$ automatically.
We set the search domain in this case by a help of human intuition 
and numerical evaluations of the target function at several points.
The second problem is to move over the singular locus of the Pfaffian system
without numerical instability.
In this case, we pose the conditions
$x_{33} \leq -0.01$,  $y_1 \leq -0.01$ and $y_2 \leq -0.001$,
because the variety $x_{33}=y_1=y_2=0$ lies in the singular locus of the Pfaffian system.

\bigbreak
\noindent
{\bf Magnetism data}

We consider the problem to 
minimize 
$$
F(x,y,1)   \exp\left( - \sum_{1 \leq i \leq  j \leq 3} S_{ij} x_{ij} 
                      - \sum_{i} S_i y_i \right)
$$
on 
\begin{eqnarray*}
& & (x_{11}, x_{12}, x_{13}, x_{22}, x_{23}, x_{33}, y_1,y_2, y_3) \\
&\in & E=[-30,30] \times  [-30,30] \times [-30,30]
                  \times [-30,30] \times [-30,30]
                                 \times [-30,-0.01] \\
 & & \quad\quad
      \times [-30,30] \times [-32,-0.001] \times [-30,32]
\end{eqnarray*}
where 
\begin{eqnarray*}
&&
  (S_{11}, S_{12}, S_{13}, 
           S_{22}, S_{23},
                   S_{33},
   S_{1}, S_{2}, S_{3}) \\
&=&
  ( 0.045,
    -0.075,
    0.014,
    0.921,
    -0.122,
     0.034,
     0.082,
    -0.959,
     0.131).
\end{eqnarray*}
The result is that
the minimum $0.4373096253840751950$ is taken at \\
$x=
x_o=\pmatrix{ 7.065 & -0.032/2 & 3.422/2 \cr
             -0.032/2 & 5.339    & 24.922/2 \cr
             3.422/2  & 24.922/2 & -13.693 \cr}$,
$ y=(1.642,\underline{-31.99},\underline{31.992})
$
with the grid size $0.01$ and the 4th order Runge-Kutta method.
Although $y_2$ and $y_3$ are on the border with this grid size,
we can observe that the change of the target value is relatively small,
when we enlarge the domain.
In fact, we started the holonomic gradient descent from the optimal point,
obtained by Wood's method \cite{wood}, \cite[{\tt toc.html}]{web},
which is \\
$x=\pmatrix{ 5.985 & 8.478/2 & 2.902/2 \cr 
             8.478/2 & 6.869   & 16.732/2 \cr
             2.902/2 & 16.732/2 & -12.853 \cr}$,
$y=(9.762,-28.770, 24.142)$.
The optimal value of the target function is $0.4421940620633763292$.
If we restart the holonomic gradient descent from the point 
$x_o$ by recalculating the integral values,
we get a new optimal point and the target value changes only about $10^{-5}$.
Since the significant figures of the given data $S_{ij}, S_i$ are $2$ digits,
we may conclude that there seems to be a variety which gives the optimal value of 
the target function.
Our method finds a point in the variety and moves in the variety.

The statistical problems considered in this section can be solved 
by a different method.
A.~T.~A.~Wood \cite{wood} expressed the Fisher-Bingham integral of the case $n=2$
as a single integral with the integrand expressed by a modified Bessel function. 
He gives a method to solve a minimization problem 
equivalent to our problem (\ref{eqn:obj}) based on this single integral representation.
We implement his method by the statistical computing system R and obtain analogous computational results with us.
The program is obtainable from \cite[{\tt toc.html}]{web}. 

Although our two statistical problems can be solved by his different method,
the advantage of our approach is that our method is a general algorithm
which can be applied to a broad class of problems,
which will be presented in forthcoming papers,  and
is based on a holonomic system of differential equations.
We note that this point of view of holonomic system
has been emphasized by few literatures in statistics.

{\it Acknowledgements}.
We thank to Prof. K.Takeda for comments on optimization methods.

\section{Appendix: Introduction to Holonomic Ideals}

Although we want to suppose people with different disciplines as readers of this paper,
the theory and algorithms for holonomic ideals are not very popular
and facts needed for the holonomic gradient descent are in diverse 
literatures.
We will present an introductory overview on 
these well-known facts of holonomic ideals and algorithms
(see \cite{SST} and its references for proofs and original articles).

We denote by $D$ the ring  of differential operators with polynomial
coefficients
$$D = {\bf C}\langle x_1, \ldots, x_d, \pd{1}, \ldots, \pd{d} \rangle, $$
which is also called the Weyl algebra.
This is an associative non-commutative ring and $x_i$ and $\pd{j}$ have
the commuting relations
$$ x_i x_j = x_j x_i,
   \pd{i} \pd{j} = \pd{j} \pd{i},
   \pd{i} x_j = x_j \pd{i} + \delta_{ij}$$
where $\delta_{ij}$ is Kronecker's delta.
Elements in $D$ are often expressed by using the multi-index notation
such as
$ x^\alpha \pd{}^\beta 
= \prod_{i=1}^d x_i^{\alpha_i} \prod_{i=1}^d \pd{i}^{\beta_i}
$.
$|\alpha|$ is defined by $\alpha_1 + \cdots +\alpha_d$.
By utilizing the commuting relations, any element of $D$ can be transformed
into the normally ordered form
$ \sum_{(\alpha,\beta) \in E} c_{\alpha\beta} x^\alpha \pd{}^\beta $.
For example, the normally ordered form of $ \pd{1} x_1 \pd{1}$ is 
$x_1 \pd{1}^2 + \pd{1}$.
Elements of $D$ acts on a function $f(x_1, \ldots, x_d)$ by
$$ x^\alpha \pd{}^\beta \bullet f
  = x^\alpha \frac{ \partial^{|\beta|} f}
                  {\partial x_1^{\beta_1} \cdots \partial x_d^{\beta_d}}
$$
where we denote by $\bullet$ the action.

Let us introduce one more important ring $R$, which we call
the ring of differential operators with rational function coefficients,
$$ R = {\bf C}(x_1, \ldots, x_d) \langle \pd{1}, \ldots, \pd{d} \rangle $$
where we denote by ${\bf C}(x_1, \ldots, x_d)$  the field of
rational functions in $x_1, \ldots, x_d$.
This is also an associative non-commutative ring and
the commuting relations are
$ \pd{i} \pd{j} = \pd{j} \pd{i}$ and
$ \pd{i} a(x) = a(x) \pd{i} + \frac{\partial a}{\partial x_i}$
for $a(x) \in {\bf C}(x_1, \ldots, x_d)$.

The theory of Gr\"obner basis (see, e.g., \cite{CLO})
can be easily generalized in $D$ and $R$ as long as orders satisfy
some conditions.
Since we do not need consider general orders, we fix the order to
the graded reverse lexicographic order $\prec$ among monomials $\pd{}^\beta$
in the sequel. 
In case of $d=2$, we have
$$ 1 \prec \pd{2} \prec \pd{1} \prec \pd{2}^2 \prec \pd{1} \pd{2} 
     \prec \pd{1}^2 \prec \cdots .
$$

Let us explain some facts about Gr\"obner bases in $R$, which 
are used in this paper.
For $f \in R$, the leading term (the initial term) with respect to $\prec$
is denoted by ${\rm in}_\prec(f)$
and we regard this element as an element in 
${\bf C}(x_1, \ldots, x_d)[\xi_1, \ldots, \xi_d]$
where $\xi_i$ and $x_j$ commute each other.
For example, when  $f=(x_1 +x_2) \pd{1}^2 \pd{2} + (x_2^4+1) \pd{2}$,
we have ${\rm in}_\prec (f) = (x_1 +x_2) \xi_1^2 \xi_2$.
We say that $a(x) \xi^\beta$ divides $b(x) \xi^{\beta'}$  when
$\beta_i \leq \beta_i'$ for all $i$.
We call the following algorithm {\it the normal form algorithm} 
({\it the division algorithm}).

\begin{algorithm} \rm \quad \rm (${\rm \tt NormalForm}(f,G)$) \\
Input: $f$, $G=\{g_1, \ldots, g_m \}$ \\
Output: The normal form $r$ (remainder) and 
quotients $q_1, \ldots, q_m$,
which satisfy the following conditions
(a) $ f = \sum_{i=1}^m q_i g_i + r $ in $R$,
(b) $f \succeq q_i g_i $, 
(c) ${\rm  in}_\prec(g_i)$ does not divide any term of $r_{|_{\pd{}\rightarrow \xi}}$ for all $i$. 
\begin{enumerate}
\item $r \leftarrow 0$, $q_i \leftarrow 0$.
\item Call ${\rm {\tt wNormalForm}}(f, G)$.
We suppose that the output is $r', q_1', \ldots, q_m'$.
\item $f \leftarrow r'-{\rm in}_\prec(r')$,
      $r \leftarrow r + {\rm in}_\prec(r')$,
      $q_i \leftarrow q_i + q_i'$.
 If  $f=0$, then return $r, q_1, \ldots, q_m$
 else goto 2.
\end{enumerate}
\end{algorithm}

\begin{algorithm} \quad \rm (${\rm \tt wNormalForm}(f,G)$) 
\begin{enumerate}
\item $r \leftarrow f$, $q_i \leftarrow 0$
\item {\it If} there exists $i$ such that ${\rm in}_\prec (g_i)$ divides
${\rm in}_\prec(r)$ {\it then} \\
\ \quad\quad\quad  $r \leftarrow r - c(x) \pd{}^\beta g_i$
where $c(x) \pd{}^\beta$ is chosen so that
${\rm in}_\prec(r) - c(x) \xi^\beta {\rm in}_\prec(g_i) = 0$; \\
\ \quad\quad\quad  $q_i \leftarrow q_i + c(x) \pd{}^\beta$; \\
{\it else} return $r, q_1, \ldots, q_m$. 
\item goto 2.
\end{enumerate}
\end{algorithm}

\begin{example} \rm \label{example:bess}
We compute the normal form of 
$f=\pd{1} \pd{2}^3$
by 
$g_1 = \underline{\pd{1} \pd{2}}+1$,
$g_2 = \underline{2 x_2 \pd{2}^2} -\pd{1}+ 3 \pd{2} +2x_1$
with the graded reverse lexicographic order.
Since we have
\begin{eqnarray*}
&& \pd{1}\pd{2}^3 - \pd{2}^2 g_1 = - \pd{2}^2 \\
&& - \pd{2}^2+\frac{1}{2 x_2} g_2 = \frac{1}{2x_2}(-\pd{1}+3 \pd{2} + 2 x_1) =: f^*,
\end{eqnarray*}
the normal form is $f^*$ and
$q_1 = \pd{2}^2$ and $q_2 = -\frac{1}{2x_1}$.
This example is taken from \cite{OST}.
\comment{
We note that the set $\{ g_1, g_2 \}$ is a system of differential operators
for the Bessel function in $2$ variables (see, e.g., \cite{OST}).
}
\end{example}

Let $I$ be a left ideal in $R$.
A finite set $G = \{ g_1, \ldots, g_m \}$, $g_i \in R$ is called 
a Gr\"obner basis of $I$ with respect to $\prec$ when
$\langle {\rm in}_\prec(g_1), \ldots, {\rm in}_\prec(g_m) \rangle
= \langle {\rm in}_\prec(f) \,|\, f \in I \rangle$.
Here, $\langle h_1, \ldots, h_m \rangle$
is the set $\sum_{i=1}^m {\bf C}(x_1, \ldots, x_d)[\xi_1, \ldots, \xi_d] h_i$,
which is the ideal generated by $h_1, \ldots, h_m$ in 
${\bf C}(x_1, \ldots, x_d) [\xi_1, \ldots, \xi_d]$.
\comment{
Although, we use $\langle, \rangle$ to specify the generators of non-commutative
rings $D$ and $R$,
we also use the notation $\langle h_1, \ldots, h_m \rangle$ 
to denote the ideal generated by $h_1, \ldots, h_m$ here.
It might be a little confusing, but the meaning will be clear in the context.
}
A Gr\"obner basis can be obtained by the Buchberger algorithm.
The proof is analogous with the case of the ring of polynomials
(see, e.g., \cite[Chapter 2]{CLO}).

Let $G$ be a Gr\"obner basis.
The element $\pd{}^\beta$ is called a standard monomial
when none of ${\rm in}_\prec(g)$, $g\in G$ divides $\xi^\beta$. 
Any normal form is a sum of standard monomials over ${\bf C}(x_1, \ldots, x_d)$.

\begin{example} \rm
This is a continuation of the previous example.
Put $g_3=\underline{\pd{1}^2}  
  - 3 \pd{1}\pd{2} - 2 x_1 \pd{1} + 2 x_2 \pd{2} -2 $.
Then, the set $\{g_1, g_2, g_3\}$ is a Gr\"obner basis of the left ideal
in $R$ generated by $g_1$ and $g_2$.
The set of the standard monomials is
$\{ 1, \pd{1}, \pd{2} \}$.
\end{example}

The output $r$ of the normal form algorithm depends on which index $i$
we choose in the step 2 in the algorithm ${\tt wNormalForm}$.
\begin{theorem} \label{theorem:unique}
Let $f$ be an element of $R$.
If $G$ is a Gr\"obner basis, then the normal form $r$ of $f$ by $G$ 
is unique. 
\end{theorem}

{\it Proof}\/.
Suppose that we have two different normal forms $r_1$ and $r_2$.
Since we have $r_1 - r_2 \in I$, ${\rm in}_\prec(r_1-r_2)$ is divisible by
an ${\rm in}_\prec (g_i)$ by the definition of Gr\"obner basis.
But it contradicts to that  $r_i$ is a sum of standard monomials
over ${\bf C}(x_1, \ldots, x_d)$. Q.E.D.

\medbreak

When the number of the standard monomials is finite,
the ideal $I$ is called {\it a zero-dimensional ideal}\/.
It follows from Theorem \ref{theorem:unique} that
the number is equal to the dimension of $R/I$ as the vector 
space over ${\bf C}(x_1, \ldots, x_d)$
(see, e.g., \cite[Chapter 5]{CLO}).
It implies that the number of the standard monomials does not depend
on Gr\"obner bases. 
The dimension is called the {\it holonomic rank} of $I$.

We call $c(x) \pd{}^\beta$, $0 \not= c(x) \in {\bf C}(x_1, \ldots, x_d)$,
a non-monic standard monomial when $\pd{}^\beta$ is a standard monomial.
Let $S=\{s_1=1,s_2, \ldots, s_p\}$ 
be a set of (independent) non-monic standard monomials of the Gr\"obner basis $G$
such that $p=\sharp S ={\rm dim}_{{\bf C}(x_1, \ldots, x_d)}\, R/RG$.
Put $Q=(s_i \bullet g \,|\, s_i \in S)^T$.
In order to apply  holonomic gradient descent, 
we need to compute the $p \times p$ matrix $P_i$ in
the Pfaffian equations
$$
 \frac{\partial Q}{\partial {x_i}} = P_i Q, \quad i=1, \dots, d.
$$
which is (\ref{eq:pfaffian}) in the main text.
To obtain the matrix $P_i$, we apply the normal form algorithm to $\pd{i} s_j$.
Then, the coefficient of the normal form of $\pd{i} s_j$
with respect to $s_k$ is the  $(j,k)$-th
element of $P_i$.
This is the step 2 of the Algorithm \ref{algorithm:hgd0} in the main text.

\begin{example} \rm
This is a continuation of the previous example.
We choose $S=\{ 1, x_1 \pd{1}, x_2 \pd{2} \}$.
Then, we obtain
$$P_1 =
\pmatrix{
0& \frac{ 1}{ {x}}& 0 \cr
 - {x}& \frac{   2   {x}^{ 2} + 1}{ {x}}&  -  2  {x} \cr
 - {y}& 0& 0 \cr
},
P_2=
\pmatrix{
0& 0& \frac{ 1}{ {y}} \cr
 - {x}& 0& 0 \cr
 - {x}& \frac{\frac{ 1}{ 2}}{ {x}}& \frac{\frac{ - 1}{ 2}}{ {y}} \cr
}
$$
where $x=x_1$ and $y=x_2$.
We can utilize several packages to perform this computation.
Among them, we use the package ``yang'' \cite{yang} 
on {\tt Risa/Asir}\footnote{\cite{asir}, {\tt http://www.math.kobe-u.ac.jp/Asir}},
because it can perform a large scale computation, 
which is required in our applications.
The code to obtain the result above is
{\footnotesize
\begin{verbatim}
import("yang.rr");
def ex1() {
 yang.define_ring([x,y]);
 L1=dx*dy+1;
 L2=dx^2-2*x*dx+2*y*dy+1;
 L3=2*y*dy^2+3*dy-dx+2*x;
 L=[L1,L2,L3];
 L=yang.util_pd_to_euler(L,[x,y]);
 L=map(nm,L);
 L=map(dp_ptod,L,[dx,dy]);
 G=yang.buchberger(L);
 S1=yang.constant(1);
 Sx=yang.operator(x);
 Sy=yang.operator(y);
 Base=[S1,Sx,Sy];
 Pf=yang.pfaffian(Base,G);
 return Pf; 
}
ex1();
\end{verbatim}
}
Since we have $\pd{1} = \frac{1}{x_1} s_2$ and $\pd{2} = \frac{1}{x_2} s_2$,
the gradient
$\nabla g = \pmatrix{ \frac{\partial g}{\partial x} \cr
                      \frac{\partial g}{\partial y} \cr}$
is equal to $A G$ where
the matrix $A=(a_{ij})$ is 
$\pmatrix{0 & \frac{1}{x_1} & 0 \cr
          0 & 0             & \frac{1}{x_2} \cr}$.
\end{example}
 

We call a function $F$ {\it a holonomic function} when it satisfies
ordinary differential equations for all variables.
In other words, $F$ satisfies
\begin{equation} \label{eq:hode}
 \sum_{k=0}^{r_i} a^i_k(x_1,\ldots, x_d) \partial_i^k \bullet F = 0, \quad
   a^i_k \in {\bf C}[x_1, \ldots, x_d], \quad i=1, \ldots, d.
\end{equation}
The set of operators in $R$ which annihilate a function $F$ is a left ideal
in $R$.
In fact, if $\ell_1 \bullet F = \ell_2 \bullet F = 0$,
then we have $(\ell_1 + \ell_2) \bullet F = 0$, and
if $\ell \bullet F = 0$, then $(h \ell) \bullet F = 0$
for all $h \in R$.
We denote the set by ${\rm Ann}_R F$.
When the function  $F$ is holonomic,
${\rm Ann}_R F$ contains ordinary differential equations (\ref{eq:hode}).
Therefore, the number of standard monomials of a Gr\"obner basis 
of ${\rm Ann}_R F$ is less than or equal to $\prod_{i=1}^d r_i$.
In other words, we have
${\rm dim}_{{\bf C}(x_1, \ldots, x_d)} \, R/{\rm Ann}_R F \leq \prod_{i=1}^d r_i$.
Conversely, we have the following theorem.
\begin{theorem} 
Let $I$ be a left ideal in $R$.
If $m:={\rm dim}_{{\bf C}(x_1, \ldots, x_d)} R/I$ is finite, then
the left ideal $I$ contains an ordinary differential operator
for any variable $x_i$.
\end{theorem}

{\it Proof}\/.
$1, \pd{i}, \pd{i}^2, \ldots, \pd{i}^m$ are linearly dependent
in $R/I$, which we regard as a vector space over ${\bf C}(x_1, \ldots, x_d)$.
This implies that there exist rational functions $c_k(x)$ such that
$\sum_{k=0}^m c_k(x) \pd{i}^k \in I$.  
Q.E.D.

\medbreak
This theorem is an analogy of the elimination theorem. 
The elimination in $R$ can be done by an analogous method in case of 
the ring of polynomials (see, e.g., \cite[Chapter 3]{CLO}).

We have worked in the ring $R$. 
If we need to consider integrals of $F$,
we need the theory and algorithms for the Weyl algebra $D$.
Let us proceed on a discussion on $D$.

We first note that we can easily generalize the Gr\"obner basis theory
for term orders $\prec$ in $D$.
For example, in case of $d=2$, 
the Gr\"obner basis theory works for the graded reverse lexicographic order
such that
$ 1 \prec x_1 \prec x_2 \prec \pd{1} \prec \pd{2} \prec x_1^2 \prec \cdots $.

We introduce the notion of a holonomic ideal.
Let $F_k$ be the set of elements in $D$ of which order is less than or equal to $k$.
In other words, $F_k$ is a ${\bf C}$-vector space
spanned by $x^\alpha \pd{}^\beta$, $|\alpha| +|\beta| \leq k$.
$\{ F_k \}$ is called the Bernstein filtration.
A left ideal $I$ in $D$ is called {\it a holonomic ideal}
when
${\rm dim}_{\bf C} F_k/F_k \cap I = O(k^d)$
for sufficiently large numbers $k$.
The quotient $D/I$ is called {\it a holonomic $D$-module} when $I$ is a holonomic ideal.
We note that the dimension agrees with the number of standard monomials
of which total degree is less than or equal to $k$
with respect to a Gr\"obner basis of $I$ by the graded reverse lexicographic
order (see, e.g., \cite[Chapter 9]{CLO}).

\begin{lemma} \label{ap:lemma:holonomicElimination}
Let $I$ be a holonomic ideal in the ring of differential operators
$D = {\bf C} \langle x_1, \ldots, x_d, \pd{1}, \ldots, \pd{d} \rangle$.
We choose a set of $d+1$ variables from the set \\
$\{ x_1, \ldots, x_d, \pd{1}, \ldots, \pd{d}  \}$ and denote it by $V$.
Then, the elimination ideal $I \cap {\bf C} \langle V \rangle$ contains a non-zero element.
\end{lemma}

{\it Proof}\/.
Consider the ${\bf C}$-linear map
$$ \rho_k \,:\, {\bf C}\langle V \rangle \cap F_k \ni \ell \mapsto [\ell] \in F_k/F_k\cap I $$
The dimension of the ${\bf C}$-vector space ${\bf C}\langle V \rangle \cap F_k$
is ${{d+1+k} \choose {d+1}} = O(k^{d+1})$.
On the other hand, we have
${\rm dim}_{\bf C}\, F_k/F_k\cap I = O(k^d)$ because $I$ is a holonomic ideal.
Since
${\rm dim}_{\bf C} \, {\rm Im}\, \rho_k =
 {\rm dim}_{\bf C} \, {\bf C}\langle V \rangle \cap F_k - 
 {\rm dim}_{\bf C} \, {\rm Ker}\, \rho_k
$,
we conclude that the vector space ${\rm Ker}\, \rho_k$ contains a non-zero element.
Q.E.D.

\medbreak
When $I$ is a holonomic ideal, the number of standard monomials is
infinite in general.  
It is natural to ask if there is a zero-dimensional ideal in $D$.
However, the following theorem claims that the holonomic ideals are
the biggest ideals and there is no zero-dimensional ideal in $D$
\begin{theorem} {\rm (Bernstein inequality)}
Let $I$ be a left ideal in $D$.
Suppose that $I \not= D$.
There exists a constant $p$ such that
${\rm dim}_{\bf C} F_k/F_k\cap I = O(k^p)$ for sufficiently large $k$
and the inequality $p \geq d$ holds.
\end{theorem}

Let us explain a relation of a holonomic ideal in $D$ and a zero dimensional
ideal in $R$.
For a left ideal $I$ in $D$,
we denote by $RI$ the left ideal in $R$ generated by elements in $I$.
It follows from the Lemma \ref{ap:lemma:holonomicElimination}
that if $I$ is a holonomic ideal,
then $I$ contains an ordinary differential operator for any variable $x_i$
and then $RI$ is a zero-dimensional ideal.
Conversely, we have the following theorem.

\begin{theorem}
If $J$ is a zero-dimensional ideal in $R$, then
$J \cap D$ is a holonomic ideal in $D$.
\end{theorem}

An elementary proof of this fact is found in the appendix of \cite{takayama1992}.
We emphasize that when we are given a set of generators of $J$,
it is not necessarily a set of generators of $J \cap D$.
The ideal $J \cap D$ is called {\it the Weyl closure} of $J$.
An algorithm to find a set of generators of the Weyl closure is given by H.~Tsai
(Algorithms for associated primes, Weyl closure, and local cohomology of $D$-modules.  
Lecture Notes in Pure and Appl. Math., 226, 169--194, 
Dekker, New York, 2002).
Although we can make a lot of constructions for $0$-dimensional ideals in $R$,
for algorithms in $D$ like $D$-module theoretic integration algorithms, 
we often require that {\it inputs are holonomic}.
However,  finding a set of generators of $J \cap D$ requires
a high complexity. 
It often makes computational bottlenecks.

\begin{example} \rm
We consider the function $f(x,y,z)={\rm exp}(1/g)$
where $g=x^3-y^2 z^2$.
The function $f$ is annihilated by  first order operators
$$ g^2 \pd{x} + 3 x^2,
   g^2 \pd{y} - 2 y z^2,
   g^2 \pd{z} - 2 y^2 z
$$
The left ideal $I$ generated by these operator is not holonomic.
The Weyl closure $J=RI \cap D$ is holonomic.
The below is a {\tt Macaulay 2}\footnote{{\tt http://www.math.uiuc.edu/Macaulay2}} 
script to check the holonomicity and 
find the Weyl closure of $RI$.
{\footnotesize
\begin{verbatim}
loadPackage "Dmodules"
D=QQ[x,y,z,dx,dy,dz, WeylAlgebra=>{x=>dx,y=>dy,z=>dz}];
I = ideal((x^3-y^2*z^2)^2*dx+3*x^2,
          (x^3-y^2*z^2)^2*dy-2*y*z^2,
          (x^3-y^2*z^2)^2*dz-2*y^2*z);
II=inw(I,{0,0,0,1,1,1});
print(dim II);  --- the output 4 implies that it is not holonomic.
J=WeylClosure I;
print(toString(J));
JJ=inw(J,{0,0,0,1,1,1});
print(dim JJ);  --- the output 3 implies that it is holonomic.
\end{verbatim}
}
\end{example}

We close this appendix with introducing the integration ideal.
The next fact is 
the fundamental fact for holonomic ideals and integrations.
\begin{theorem}
If $I$ is a holonomic ideal, then the integration ideal
$(I + \pd{d} D) \cap D_{d-1}$ is a holonomic ideal in $D_{d-1}$.
Here 
$D_{d-1}={\bf C}\langle x_1, \ldots, x_{d-1}, \pd{1}, \ldots, \pd{d-1} \rangle$.
\end{theorem} 
This theorem follows from the fact 
``if $D/I$ is a holonomic $D$-module, then $D/(I + \pd{d} D)$ is a holonomic
$D_{d-1}$-module''.
As to a proof of this fact, see, e.g., the Chapter 1 of the book
``J.~E.~Bj\"ork,
{\it Rings of Differential Operators}.
North-Holland, New York, 1979''.

Oaku's algorithm \cite{oaku}
to find integration ideals is explained in the Chapter 5 
of \cite{SST} in a form relevant to our applications.
We note that integration algorithms (\cite{NN}, \cite{oaku})
in $D$ use non-term orders (see,  e.g., \cite[Chapter 1]{SST}).
Modifications of this algorithm \cite{NN} 
is used in the step 1 of our Algorithm 
\ref{algorithm:hgdi}.

\begin{example} \rm
Put $f(x,t) = {\rm exp}(xt-t^3)$.
The function $f$ is annihilated by the operators
$ \pd{t} -( x-3t^2)$, $ \pd{x} -t $,
which generate a holonomic ideal $L$.
This is a {\tt Risa/Asir} code to find the integration ideal 
$(L+\pd{t} {\bf C}\langle x,t,\pd{x},\pd{t}\rangle) \cap {\bf C}\langle x, \pd{x} \rangle$.
{\footnotesize
\begin{verbatim}
import("nk_restriction.rr");
def step1() {
 L=[dt-(x-3*t^2),
   dx-t];
 I=nk_restriction.integration_ideal(L,[t,x],[dt,dx],[1,0] |  inhomo=1);
 return I;
}
step1();
\end{verbatim}
}
\end{example}

We write this introductory exposition with a few overlaps with \cite{SST}. 
For other fundamental facts, 
please refer to \cite{SST} and its references.


\bigbreak
\rightline{ Tomonari Sei,  Akimichi Takemura ${}^\sharp$}
\rightline{Department of Mathematical Informatics}
\rightline{Graduate School of Information Science and Technolgy, University of Tokyo}
\rightline{Bunkyo, Tokyo, 113-0033, Japan}

\medskip

\rightline{ Nobuki Takayama\footnote{Supported by Kakenhi 19204008}${}^\sharp$   
({\tt takayama@math.kobe-u.ac.jp}),}
\rightline{ Hiromasa Nakayama${}^\sharp$,  Kenta Nishiyama${}^\sharp$, 
 Masayuki Noro${}^\sharp$\footnote{ Authors with ${}^\sharp$ belong to the JST crest Hibi project.} }
\rightline{Department of Mathematics, Kobe University }
\rightline{Rokko, Kobe, 657-8501, Japan}

\medskip

\rightline{ Katsuyoshi Ohara\footnote{Supported by Kakenhi 22540179}} 
\rightline{Faculty of Mathematics and Physics, Kanazawa University}
\rightline{Kakuma-machi, Kanazawa, 920-1192, Japan}


\begin{thebibliography}{99}
\bibitem{yang} K.~Ohara, the yang package, 
which is included in the library of Risa/Asir.
\end{thebibliography}
\end{document}